\documentclass[a4paper,12pt, twocolumn]{article}
\usepackage[english]{babel}
\usepackage{graphicx}
\usepackage{amssymb}
\usepackage{amsmath}
\usepackage{geometry}
\usepackage{hyperref}
\hypersetup{
colorlinks=true,
linkcolor=blue,
citecolor=blue,
urlcolor=black
}

\begin{document}

\renewcommand{\refname}{References}

\title{\vspace{-1.8cm}
Toward modern educational IT-ecosystems: from learning management systems to digital platforms\thanks{The research is partially supported by the Russian Foundation for Basic Research (project 18-29-03100) and the the RF Presidential scholarship program (No. 538.2018.5).}
}

\author{
A.\,K.~Gorshenin\textsuperscript{1}}

\date{}

\maketitle

\footnotetext[1]{Institute of Informatics Problems, Federal Research
Center ''Computer Science and Control'' of Russian Academy of
Sciences, Russia; \url{agorshenin@frccsc.ru}}

\maketitle

\begin{abstract}
The development of a learning management system (LMS) as a key service seems to be very effective for creation of educational digital platforms. Such platforms for both higher education institutions and various companies can provide the opportunities for networked forms of educational communication, improve the quality of the perception of innovative technologies and support tools for progress of talented youth as well as knowledge transfer. An example of such LMS is presented. The paper focuses on the demand for further development of learning management systems, their integration with modern digital platforms and potential exploitation as key services of such platforms in the context of the current educational trends of Industry 4.0 and the global trend towards a transition to a digital economy. The implementation of artificial intelligence technologies into the educational process is mentioned as an innovative way to form IT-ecosystems of modern education.
\end{abstract}

\section{Introduction}
The progress in information technology has a significant impact on various areas of human activity, including the formation of new educational requirements and standards in the transition to the digital economy. Examples include the system of electronic school diaries, automated tools for assessing knowledge and controlling progress in studies in universities, solutions for training personnel in commercial companies, etc. The electronic forms of learning provoke increased interest among students due to implementation of familiar technologies or obtaining new user's experience. Teachers have the opportunity to demonstrate various aspects of the course interactively and automate the control of the educational process, in particular, using different types of student's testing. These aspects are important elements of the modern educational process. The development of such approaches is necessary due to the new technologies of the digital economy in all spheres, because the corresponding user competencies should be formed for the effective applications. Therefore, the development of domestic innovative educational platforms are required.

When developing electronic solutions, there are problems of the effective implementation of interaction between educational systems and users, the rapid accumulation of additional information for further knowledge extraction and obtaining analytical materials. To automate the modern educational process, various learning management systems (LMS) are implemented for general administration, forming reports for educational courses and curricula, coordination of interaction between teachers and students, monitoring of performance indicators, etc.

It should be noted (see, for example, paper~\cite{1}) that the users of learning management systems often prefer the possibility of effective communication, not the implemented educational interactive solutions. When developing learning management systems, it is necessary to take into account the quality of a system, implemented services and its content~\cite{2}, because these aspects affect a learner's satisfaction in e-Learning (see, for example, paper~\cite{3}) and therefore the success of implemented e-Learning information technologies. However, due to differences in the interpretation and evaluation of the significance of these aspects in each specific case, organizations prefer to develop their own learning management systems (see, for example, the paper~\cite{4} that presents an analysis of 113 European institutions), in particular for the implementation of adaptive solutions~\cite{5}.

The creation of universal interdisciplinary and superorganizational solutions could be very perspective. These systems also could provide some special tools (for example, remote access to scientific equipment for laboratory work~\cite{6} or modern research support IT-systems~\cite{7}) for the widest possible coverage of the demands of different courses.

One of possible implementations of the automated testing service was proposed in the papers~\cite{8, 9} within the framework of a client-server paradigm. It could be an actual component of the innovative educational digital platform. However, the drawbacks of a client-server paradigm are well known, therefore, in order to meet the requirements for the development of modern information systems the cloud solutions for a learning management system should be used (for example, based on Software as a Service (SaaS) model). It implies the thin clients for students and teachers in order to use system's functional capabilities.

\begin{figure*}[h]
\centerline{\includegraphics[width=\textwidth,height=0.5\textheight]{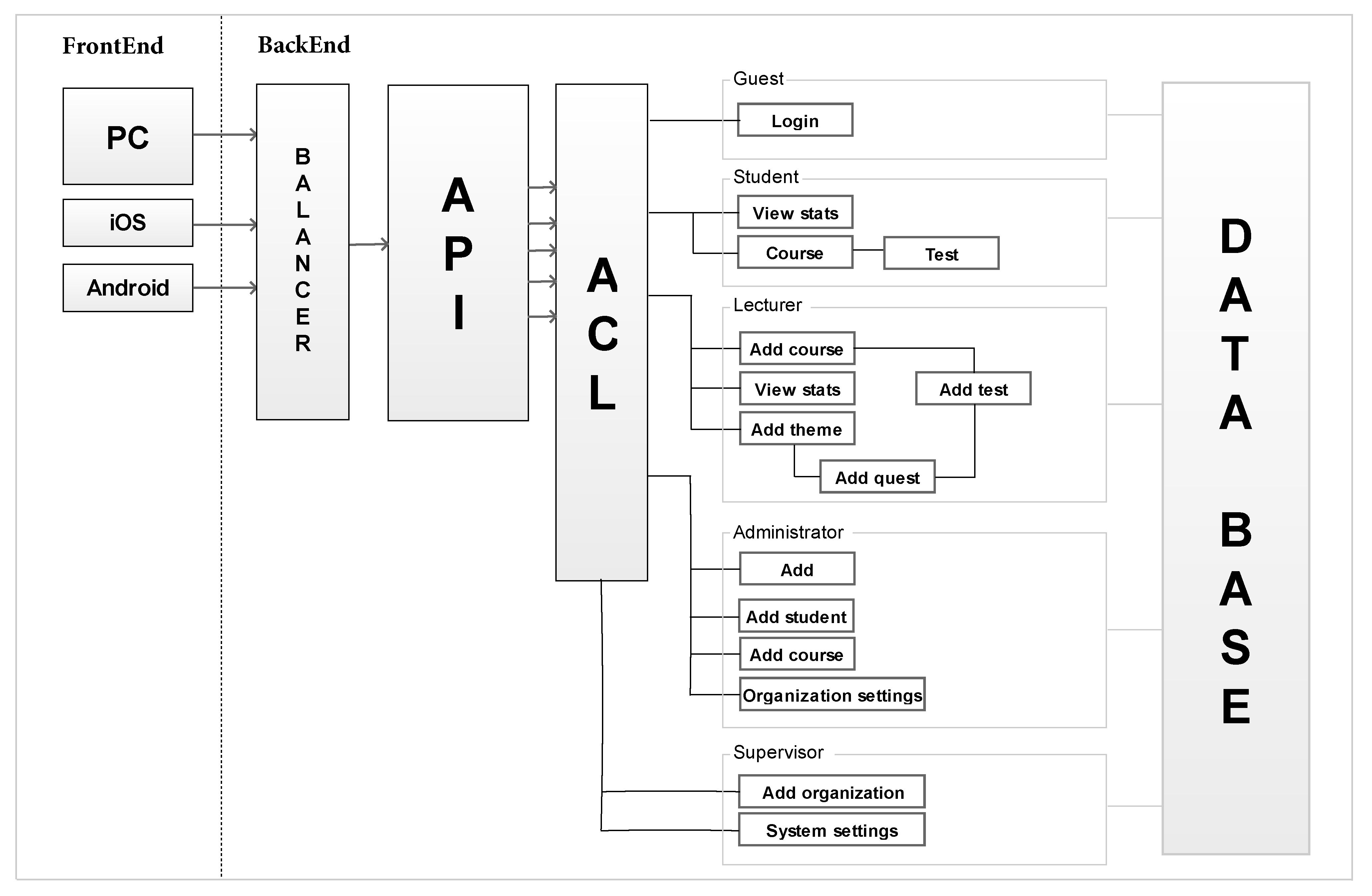}}
\caption{Architecture of the LMS.}
\label{FigArchitecture}
\end{figure*}

The application of cloud-based technologies in education (see, for example, paper~\cite{10}) should be noted, because it allows solving a number of technical problems, in particular, it is possible to reduce infrastructure costs and simplify the scaling problem. The use of mobile apps for the popular platforms as thin clients makes possible to attract a target audience focused on innovative digital technologies as well as leads to a positive effect in learning (see, for example, paper~\cite{11}). In addition, there is an opportunity to use built-in technical solutions~\cite{12}: cameras, fingerprint scanners, etc., which can be used to pass the authentication procedure or reduce some third-party vulnerabilities. These are very important problems for distance- and e-Learning.

Some specific issues can be integrated into LMS (or digital platform). For example, modern data mining and statistical data processing techniques (see, for example, papers~\cite{13,14,15,16}) can be very perspective including the framework of the modern educational process.

The paper focuses on the demand for further development of LMS, their integration with modern digital platforms and potential exploitation as key services of such platforms in the context of the current educational trends of Industry 4.0 and the global trend towards a transition to a digital economy.

\section{Digital platforms within digital economy framework}

The basis for the digital economy is data in digital form, innovative principles of data processing (in wide sense) lead to an information environment, taking into account the needs of citizens and society, as well as the new technology industry.

One of the key components of the digital economy are platforms and technologies that create competencies for the development of markets and industries, as well as a new environment for the effective interaction of various actors. Digital platforms provide a unified information environment with a help of innovative IT-solutions to reduce transaction costs. The problem of analysis, optimization and restructuring of relations between participants is greatly simplified. Such platforms allow creating new products and services, forming ecosystems. It should be noted that this approach fully corresponds to the ideology of Industry 4.0~\cite{17} both from the point of view of business, and from the standpoint of education and science.

Indeed, the world scientific community follows to a new paradigm for scientific research: significant scientific results can be obtained only with help of the analysis of large data sets (more often that is Big Data) accumulated in subject areas. For their analytical processing, it is necessary to create new methods and teach specialists with a new set of competencies, in particular, within the engineering professions.

\section{LMS as a key service of educational digital platforms}
Development of a key (and relevance) service attracts to the digital platform many others ones. For educational digital platforms, the development of an LMS as a key service seems to be very effective. In this section, we discuss the basic requirements for such LMS, as well as a certain background created by the author in this area~\cite{18, 19}.

One of the possible implementations of the automated tested service for electronic support of the educational process within the framework of a client-server paradigm had been proposed in paper~\cite{9}. It provides the following advantages: there is no duplicating code on the client side, moderate requirements for user hard- and software, increasing security due to the organization of access policies on the server. However, the drawbacks are also well known: the dependence of the system's performance and availability on the state of the server, the complexity of administration, expensive server hardware, and increasing Internet traffic under a complication of the client applications.

Modern learning management systems should provide scalability, fault tolerance, etc. The effective solutions can be built on a cloud platform with a SaaS model. It implies the thin clients for students and teachers in order to use system's functional capabilities. Implementation of cloud solutions can significantly reduce the infrastructure costs. Moreover, the system becomes more flexible concerning demands for computational capability. Therefore, the cloud solutions in the educational problems could be very useful.

\subsection{Architecture of LMS}

The main architecture requirements for the proposed cross-platform learning management system based on cloud technologies with the implementation of mobile client apps are as follows:
\begin{itemize}
\item The system should be based on cloud technologies to provide simultaneous access for a large number of users.
\item The system needs to be cross-platform and support mobile apps.
\item The system should be universal enough for educational services both in higher education institutions and in companies.
\item All data, including learning materials, tests, correct answers, etc., are stored on the server. Access is provided exclusively in secure mode.
\item Each user has a unique personal account with the implementation of different levels of access rights depending on the role in the system.
\item The administrator can add authorized agents of educational institutions or companies and provides the correctness of system's work.
\item The authorized agent of the organization can change information about the company (for example, its name, list of subjects, schedule, etc.), add teachers (lecturers), subjects and groups, delegate access rights, etc.
\item Teachers can create and edit learning and test materials, track the completion of homework. Tests can be created on the basis of the course sections, in particular without reference to any particular topic. There should be an option of "random" formation of questions from the given sections and the order of the withdrawal of possible answers for them.
\item Students should be able to find educational materials on any accessible subject, perform homework, pass tests, have a feedback from teachers and other students within the system in their personal account.
\end{itemize}

The architecture~\cite{18} of the cross-platform learning management system, including backend and frontend, is presented in Fig.~\ref{FigArchitecture}. 

\subsection{Brief description of LMS}

\begin{figure}[h]
\centerline{\includegraphics[width=0.35\textwidth,height=0.47\textheight]{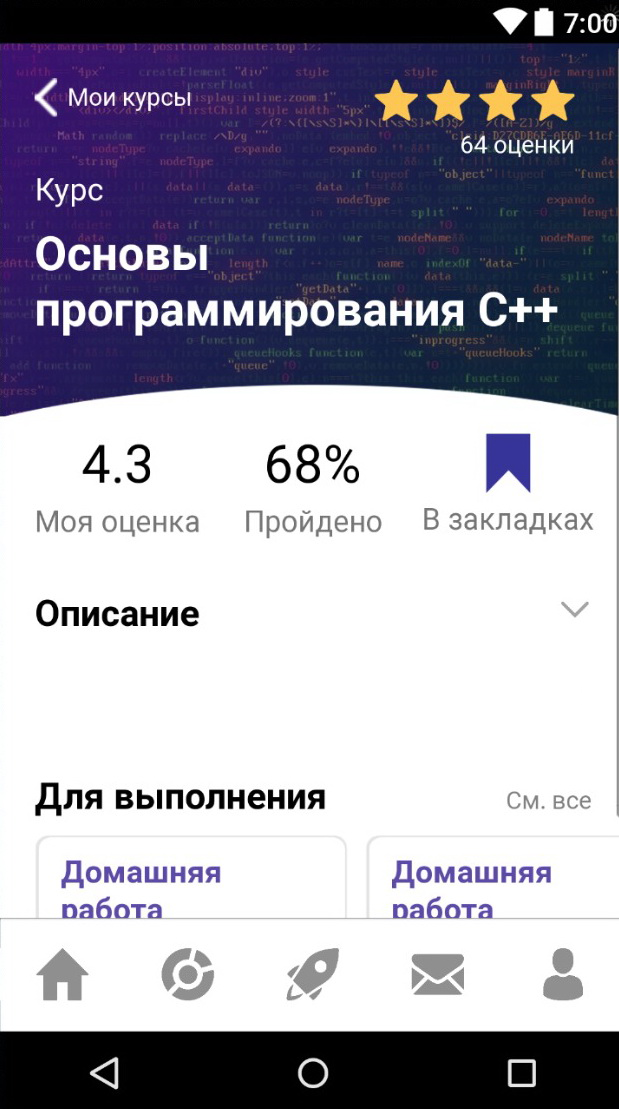}}
\caption{Design of mobile app, Android OS (Russian language interface).}
\label{FigAndroidApp}
\end{figure}

\begin{figure*}[h]
\centerline{\includegraphics[width=0.65\textwidth,height=0.47\textheight]{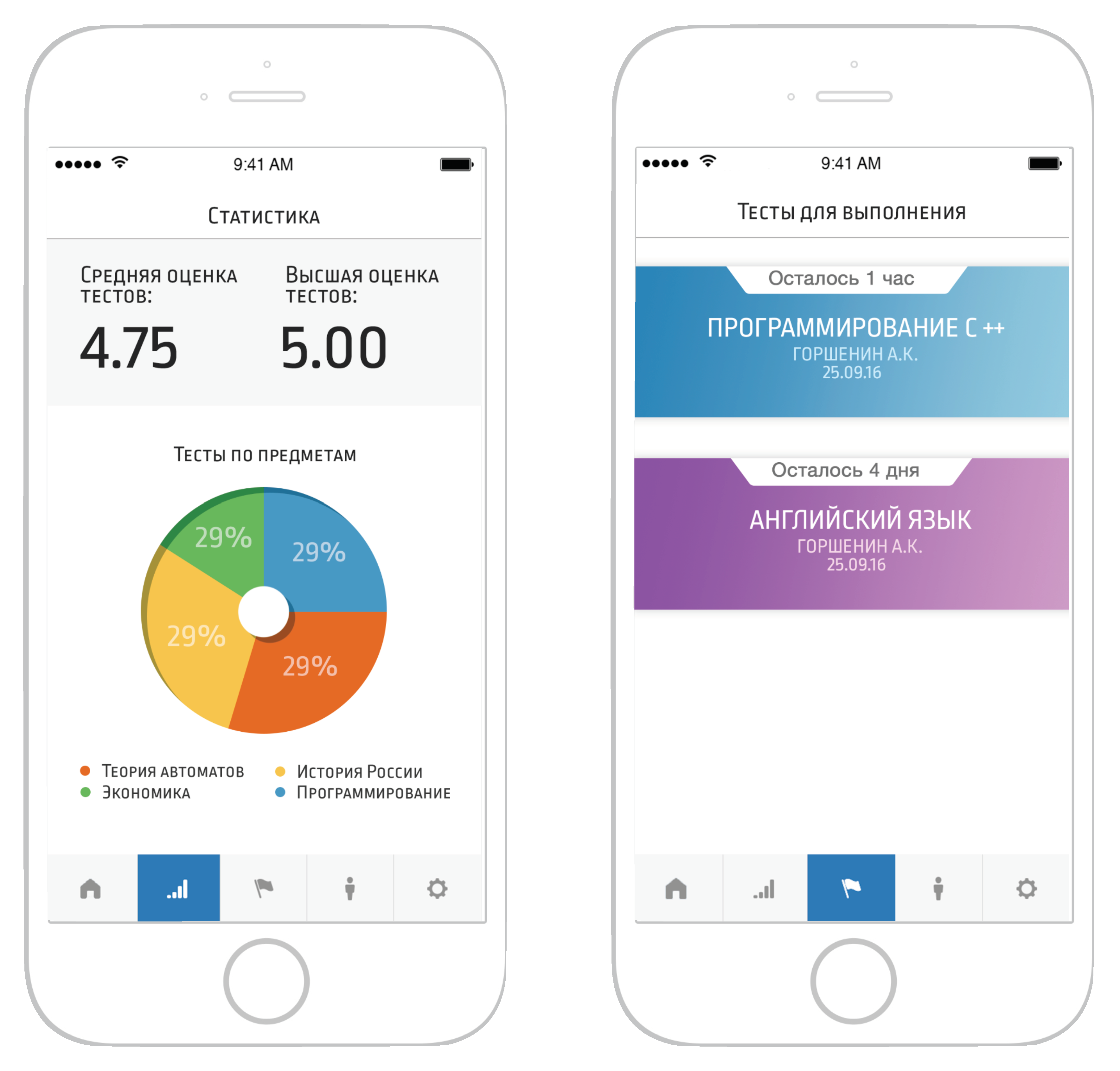}}
\caption{Design of mobile app, iOS (Russian language interface).}
\label{FigiOSApp}
\end{figure*}

In this section, examples of the design of mobile apps for Android (see Fig.~\ref{FigAndroidApp}) and iOS (see Fig.~\ref{FigiOSApp}) are discussed, and the results of testing in student focus groups with help of the developed LMS~\cite{18,19} are demonstrated. The statistics of progress, points in subjects, and the proportion of completed tasks are displayed. Students can be ranked by grading, circular and bar charts to analyze students' progress, notifications about events occurring in the system, messages, etc., are supported. The use of the Russian language interface both in applications and in the desktop version is determined by the orientation on the target audience in Russia. If necessary the interfaces of apps and the desktop version can be modified for any language by implementing the corresponding style files.

It should be noted that technologies for messaging are very important for the further transition to the digital platform (for example, as an element of educational social networks). Within the application interface, the scores received by the user for different tests are demonstrated, the percentage of completed tasks is specified (for example, in case when several trials or student's breaks are allowed), and the remaining time is indicated. The pie chart shows the various courses and the corresponding percentages of completed tasks.

The trial sets of questions on computer science and economics were suggested initially for students of MIREA -- Russian Technological University and National Research University Higher School of Economics~\cite{19}. The task for focus groups was not to make the test as fast or successful as possible. The most interesting issue was the complex interaction with the system, the evaluation of user experience and the convenience of the interface. Based on the results of testing, students were interviewed about the satisfaction with the quality of the LMS. In particular, a significant interest was noted in this form of testing and  interaction with teachers. All students would participate in the re-testing, including increasing the number of disciplines.

\subsection{From LMS to digital platforms and educational ecosystems}

\begin{figure}[h]
\centerline{\includegraphics[width=0.32\textwidth,height=0.47\textheight]{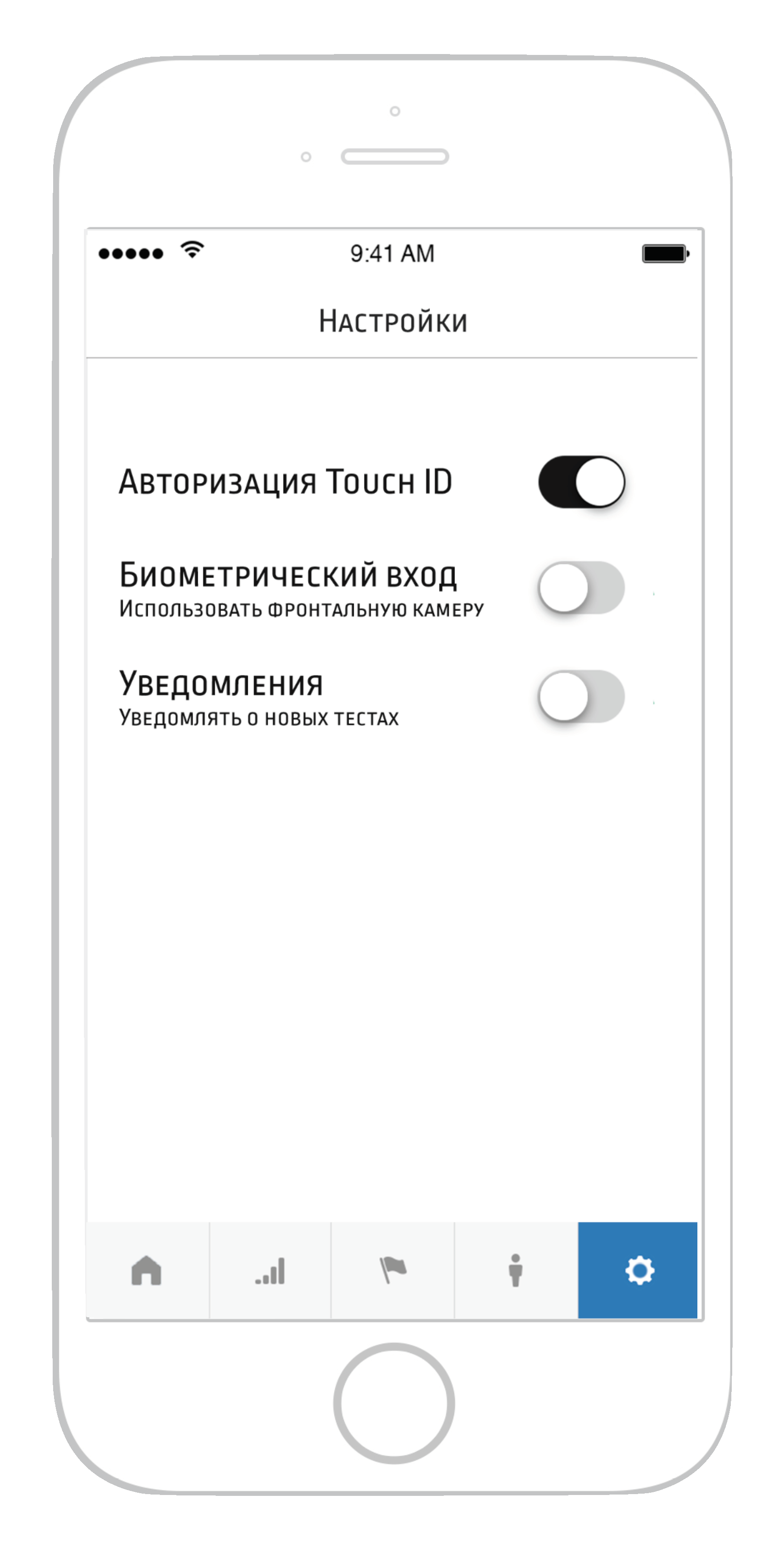}}
\caption{Possible security options.}
\label{FigBiometricsApp}
\end{figure}

A crucial part of the transition from LMS as a service to a digital platform is to provide security for the user data by using biometric solutions (see, for example, paper~\cite{20}). For example, they can be based on Apple's Face ID or Touch ID technologies blue (see sliders on Fig.~\ref{FigBiometricsApp}) and their analogs for various platforms as well as more complex solutions with 3D models for user identification. In the business segment, Sberbank successfully implements such solutions within a biometrics platform involving face, voice and retina identification.

It is important to introduce artificial intelligence technologies into the educational process. Nowadays, there are solutions for choosing individual methods for the effective e-Learning~\cite{21}, recognizing a learner's affective state and activating an appropriate response based on integrated pedagogical models~\cite{Moridis2009} including  personalized recommendations~\cite{Baylari2009}, planning demands for educational courses~\cite{22} and modelling of a learner's satisfaction~\cite{Guo2010}, evaluating a student's understanding of a particular topic of study using concept maps~\cite{Jain2014} and self-learning~\cite{23}. However, within the scope of the capabilities of the LMS mentioned in this section, artificial intelligence can directly interact both with students (for tracking their satisfaction with the learning process, solving different kinds of problems, from educational to everyday items) and with teachers (as a personal intelligent assistant). Creation, development and implementation of such approaches allow forming innovative IT-ecosystems of modern education.

\section{Conclusion}

The modern cross-platform LMS under the key global requirements is developed. Such LMS could be a key service for a high-tech domestic educational platform for both higher education institutions and various companies, providing the opportunities for networked forms of educational communication, improving the quality of the perception of innovative technologies and supporting tools for the progress of talented youth as well as knowledge transfer. The implementation of artificial intelligence technologies into the educational process is mentioned as a way to form innovative IT-ecosystems of modern education.  The described approaches fully correspond to the global trends in this field.

\section*{Acknowledgment}

The author thanks his students E.~Danilovich and D.~Khromov for preparing figures for the paper.

\end{document}